\documentclass[aps,prl,twocolumn,showpacs,superscriptaddress,groupedaddress]{revtex4-1}  % for review and submission
\usepackage{graphicx}  % needed for figures
\usepackage{dcolumn}   % needed for some tables
\usepackage{bm}        % for math
\usepackage{amssymb}   % for math
\usepackage{amsmath}   % for math
\usepackage[british]{babel}
\usepackage{hyperref}
\usepackage{aas_macros}
\usepackage[usenames]{color}

\makeatletter
\renewcommand\Dated@name{}
\makeatother

\newcommand{\appropto}{\mathrel{\vcenter{
  \offinterlineskip\halign{\hfil$##$\cr
    \propto\cr\noalign{\kern2pt}\sim\cr\noalign{\kern-2pt}}}}}

\begin{document}
\title{Cosmic Tsunamis in Modified Gravity:
Disruption of Screening Mechanisms \\from Scalar Waves}
\author{R.~Hagala$^{1}$}
\email{robert.hagala@astro.uio.no}
\author{C.~Llinares$^2$}
\author{D.F.~Mota$^1$ }
\affiliation{$^1$Institute of Theoretical Astrophysics, University of Oslo, PO Box 1029 Blindern, 0315 Oslo, Norway \\ $^2$Institute for Computational Cosmology, Department of Physics, Durham University, Durham DH1 3LE, United Kingdom}
\date{Received 18 July 2016; revised manuscript received 24 November 2016; published 10 March 2017}
\begin{abstract}
Extending general relativity by adding extra degrees of freedom is a popular approach for explaining 
the accelerated expansion of the Universe and to build high energy completions of the theory of gravity. 
The presence of such new degrees of freedom is, however, tightly constrained from several
 observations and experiments that aim to test general relativity  in a wide range of scales. 
The viability of a given modified theory of gravity, therefore, strongly depends on the existence of a 
screening mechanism that suppresses the extra degrees of freedom. 
We perform simulations, and find that waves propagating in the new degrees of freedom 
can significantly impact the efficiency of some screening mechanisms, thereby 
threatening the viability of these modified gravity theories. 
Specifically, we show that the waves produced in the symmetron model can 
increase the amplitude of the fifth force and the parametrized post-Newtonian parameters by several orders of magnitude. 
\end{abstract}
%
%\keywords{Cosmology: theory - Cosmology: miscellaneous - large-scale structure of Universe - dark energy - Methods: numerical}
\doi{10.1103/PhysRevLett.118.101301}
\maketitle
%

%
%%%%%%%%%%%%%%%%%%%%%%%%%%%%%%%%%%%%%%%%%%%%%%%%%%%%%%%%%%%%%%%%%%%%%%%%%%%%%%%%
%%%%%%%%%%%%%%%%%%%%%%%%%%%%%%%%%%%%%%%%%%%%%%%%%%%%%%%%%%%%%%%%%%%%%%%%%%%%%%%%
%%\section{Introduction}

Since 1998, it has been known that the Universe expands at an accelerating rate that is consistent with the existence of a cosmological constant \cite{1998AJ....116.1009R}. Attempts to interpret the cosmological constant as the vacuum energy from particle physics yields a mismatch of several orders of magnitude. This is known as the cosmological constant problem \cite{1989RvMP...61....1W}. A possible solution to this problem may lie on an extension of general relativity in such a way that a new gravity degree of freedom drives the accelerated expansion on large scales \cite{2012PhR...513....1C}. 

General relativity is, however, one of the most successfully tested theories in a wide range of scales; including table top experiments on Earth, laser ranging and radio wave bending in the Solar System, the rotation of black hole binaries, and the timing of pulsars \cite{2009aosp.conf..203R}. Therefore,  any modification to Einstein's gravity must include a screening mechanism to hide the new extra degree of freedom and reduce the theory to general relativity in those well tested regimes \cite{Joyce:2014kja}.

The common feature to all the screening mechanisms proposed in the literature is that they are built, and their efficiency tested, assuming the so called quasistatic approximation for the field equations.  For instance,  in scalar-tensor theories, a scalar degree of freedom is introduced into the standard Einstein-Hilbert action. This field follows the Klein-Gordon equation of motion, which determines both its time and spatial variations. When constructing screening mechanisms to hide the scalar field within the accurately tested regimes, the quasistatic approximation is invariably applied to the equations of motion for the field. This simplifies the calculations by implying that the scalar field is at rest in the minimum of the local effective potential at all points in space and time. This reduces the equation of motion to a Poisson-like equation, which is readily solved to find the approximated scalar field value at any point.
 
Notice, however, that the full equation of motion for the scalar field is, in fact, a second order differential equation in time, more similar to a wave equation. Therefore, ignoring the time evolution of the field, via the quasistatic approximation, is to shortfall effects that are only possible to realize when considering the full equation of motion \cite{2013PhRvL.110p1101L,Lindroos}. 

In this Letter, we find that, when relaxing the quasistatic approximation, the presence of waves may result in striking consequences for the efficiency and viability of the screening mechanism.  In particular, we show that energetic waves in the extra degree of freedom strongly weaken the screening process for a theory with a standard kinetic term. Therefore, modified gravity theories previously considered viable may, in fact, be ruled out by the present days gravity experiments and observational data.  To understand the implications of these waves in greater detail, we simulate a scalar degree of freedom with externally generated waves. The waves propagate radially in towards a spherically symmetric matter distribution, modeled after the Milky Way halo.

%%%%%%%%%%%%%%%%%%%%%%%%%%%%%%%%%%%%%%%%%%%%%%%%%%%%%%%%%%%%%%%%%%%%%%%%%%%%%%%%
%\subsection{The symmetron}
\emph{The Model}.---As a working example, we implement a specific form of modified gravity called the \emph{symmetron}  \cite{2010PhRvL.104w1301H}. This is a scalar-tensor theory with a  screening mechanism, constructed to hide modifications to general relativity in high density regions. In spite of this specificity, the results presented in this Letter should be considered for any modified gravity theories that have extra degrees of freedom with wave-type equations of motion. Examples of such screening mechanisms include the chameleon \cite{cham},  disformal \cite{dis4},  Dirac-Born-Infeld fields \cite{Burrage}, or $K$-mouflage \cite{Babi}.

We consider the following general scalar-tensor action for canonical scalar fields:
\begin{equation}
S=\!\!\intop \!\left[\sqrt{\!-g}\left(\frac{R}{16\pi G}-\frac{1}{2} \phi^{,\mu}\phi_{,\mu}-V\!\left(\phi\right)\right)+\sqrt{\!-\tilde{g}}\tilde{\mathcal{L}}_{m}\right]\!\mathrm{d}^{4}x,\label{eq:action}
\end{equation}
where $g$ is the Einstein frame (geometric) metric, and $\tilde{g}$ is the Jordan frame metric --- the metric dictating the geodesics of particles. $R$ is the Ricci scalar, and $\tilde{\mathcal{L}}_{m}$ is the Lagrangian density of matter (computed using the Jordan frame metric $\tilde{g}$). The field potential $V\left(\phi\right)$ is the quartic symmetron potential with the three free parameters $\mu$, $\lambda$, and $V_{0}$
\begin{equation}
V\left(\phi\right)=-\frac{1}{2}\mu^{2}\phi^{2}+\frac{1}{4}\lambda\phi^{4}+V_{0}.
\end{equation}
The Jordan frame metric $\tilde{g}$ is related to the Einstein frame metric according to the conformal transformation $\tilde{g}_{\mu\nu}=C\left(\phi\right)g_{\mu\nu}$.  The specific form of $C$ for the symmetron is $ {C\left(\phi\right)  = 1+\left(\phi / M\right)^{2}.}$ The mass scale $M$ is a free parameter that gives the strength of the interaction with the matter fields.

The equation of motion for the scalar field is
\begin{equation}
\ddot{\phi}+3H\dot{\phi}-\frac{1}{a^{2}}\nabla^{2}\phi=
-\rho\frac{C_{,\phi}\left(\phi\right)}{2C\left(\phi\right)} - V_{,\phi}\left(\phi\right),
\label{eq:eom}
\end{equation}
where a dot represents a partial derivative with respect to cosmic time,  $H=\frac{\dot{a}}{a}$ is the Hubble parameter, and $a$ is the scale factor. The Einstein frame metric is assumed to be a flat Friedmann-Lema\^itre-Robertson-Walker metric with a single scalar perturbation $\Psi_E$, specifically
\begin{equation}
\mathrm{d}s^{2}=-\left(1+2\Psi_E\right)\mathrm{d}t^{2}+a^{2}\left(t\right)\left(1-2\Psi_E\right)\mathrm{d}r^{2}.
\end{equation}
 
For convenience, we normalize the field to the vacuum expectation value of the symmetron field, $\phi_{0}\equiv\frac{\mu}{\sqrt{\lambda}}.$ As such, the new dimensionless field $\chi=\phi/\phi_{0}$ should behave in a controlled way, with $|\chi| \lesssim 1$. Also, for numerical convenience, we introduce the parameter $a_{\mathrm{SSB}}$, which defines the expansion factor at the time of spontaneous symmetry breaking.  We also introduce a dimensionless symmetron coupling constant $\beta\equiv\frac{\phi_{0}M_{\mathrm{Pl}}}{M^{2}},$ and the range of the symmetron field in vacuum, $\lambda_{0}\equiv\frac{1}{\sqrt{2}\mu}$. By taking into account these definitions, we can rewrite Eq. \eqref{eq:eom} as
\begin{equation}
\ddot{\chi}+3H\dot{\chi}-\frac{\nabla^{2}\chi}{a^2} = 
-\frac{1}{2\lambda_{0}^{2}}\left[\left(\frac{a^3_\mathrm{SSB}}{C} \frac{\rho}{\rho_0}-1\right) \chi + \chi^{3}\right],
\label{eq:eomf}
\end{equation}
where  $\rho$ is the total matter density, and $\rho_0$ the background density of the Universe.

As a working example, we fix the symmetron parameters such that $\beta = 1$, $a_\mathrm{SSB} = 0.5$, and $\lambda_0 = 0.25\,\mathrm {Mpc}/h$. 
This is equivalent to a symmetron mass 
\\$M=3.4\times10^{-4}\,M_\mathrm{pl} = 4.2\times10^{15}\,\mathrm{GeV}/c^2$.
These parameters are being widely assumed to represent a viable model that evades all the bounds from both Solar System and astrophysical data. With this choice of parameters, we aim to prove that even such a model may, in fact, be ruled out when one fully integrates the equations of motion of the field without the quasistatic assumption and, thereby, allow for the effects of the scalar waves.

%%%%%%%%%%%%%%%%%%%%%%%%%%%%%%%%%%%%%%%%%%%%%%%%%%%%%%%%%%%%%%%%%%%%%%%%%%%%%%%%
\emph{Solar System constraints}.---In order to test how screening mechanisms work in the Solar System, the community generally chooses a static, spherically symmetric matter distribution to mimic the Galaxy. We follow this approach and choose the Navarro-Frenk-White (NFW) density profile with the characteristics to represent the Milky Way Galaxy, specifically with a virial radius of $r_\mathrm{vir} = 137\,\mathrm{kpc}/h$ and concentration $c = 28$, resulting in a halo mass of $1.0\times10^{12} M_\odot$ and a circular velocity of 220~km/s at 8~kpc. The reason for the high value of the concentration is simply that we are modeling not only dark matter, but the total matter of the Milky Way, which is more concentrated than the pure dark matter halo. 
We also did the calculations with an Einasto profile with identical virial mass, 
and found that the results presented in this Letter are not very sensitive to the choice of distribution.
Because of limitations of spherical symmetry, we did not model a galactic disc.

One of the most precisely measured gravity parameters to probe deviations from general relativity is the parametrized post-Newtonian (PPN) parameter $\gamma$. It can be expressed as the ratio of the metric perturbations in the Jordan frame, $\Psi_J$ and $\Phi_J$. We find the expression for $\gamma - 1$ to be
%\[
\begin{equation}
\gamma-1=-\frac{\phi^{2}}{M^{2}}\frac{2}{\frac{\phi^{2}}{M^{2}}-2\Psi_{E}-2\Psi_{E}\frac{\phi^{2}}{M^{2}}}.
\end{equation}
%\]
In general relativity, $\gamma  = 1$ exactly. The strongest constraint to date, measured by the Cassini spacecraft \cite{Cassini}, is
$\gamma - 1 = \left( 2.1 \pm 2.3 \right ) \times 10^{-5}.$

The screening mechanism of the symmetron model works by modifying the effective potential such that the field value is pushed towards zero in high density regions -- like the inner regions of the Galaxy. This results in $\gamma - 1 \rightarrow 0$, such that the deviations from general relativity in the proximity of the Solar System are small.  The same occurs for the fifth force $F_{\phi}$ associated to the scalar field.

We calculate the $\gamma$ parameter arising from the smoothed matter distribution of the Milky Way. Note that, by using this method, we find an upper bound on the actual value of $|\gamma - 1|$ in the inner Solar System. This is because we do not include the presence of massive bodies like the Sun, which will increase the screening to some degree.
Nevertheless, most of the screening is believed to come from the matter distribution of the Galaxy because, in the symmetron model, the Solar System cannot screen itself in vacuum, and therefore, the theory depends on a working screening from the Galaxy.

%%%%%%%%%%%%%%%%%%%%%%%%%%%%%%%%%%%%%%%%%%%%%%%%%%%%%%%%%%%%%%%%%%%%%%%%%%%%%%%%
\emph{Simulations}.---Since the equation of motion is a hyperbolic partial differential equation, it can be solved as an initial value and boundary condition problem. The initial condition at $t=0$ is chosen to be the static solution of the nonlinear Klein-Gordon equation of motion. With a constant boundary condition, this would imply that the field will stay at rest forever. The boundary condition at the edge of the simulation at $r_\mathrm{max}$ is chosen to emulate incoming sinusoidal waves in the scalar field, specifically $\chi(r_\mathrm{max}, t) = \chi_0 (r_\mathrm{max}) + A\sin\left( \omega t \right)$. Possible sources of such waves will be discussed later.

We set up a radial grid, divided into linearly spaced steps $\Delta r$ up to $r_\mathrm{max} = 4\,\mathrm{Mpc}/h$. On each of the grid points we specify the matter density according to the NFW halo.  Starting from the initial value and with the inclusion of incoming waves, we evolve Eq. \eqref{eq:eomf} forward in time steps of size $\Delta t$, using the leapfrog algorithm for time integration in each grid point. Tests of this technique applied to the symmetron are presented in \cite{2013PhRvL.110p1101L,2014PhRvD..89h4023L}.
We are only interested in events that happen during the last few megayears of cosmic time, meaning that we take the approximations $z\approx0$ and $a\approx1$ in all computations.  Spatial derivatives are found using a finite difference method in spherical coordinates, assuming all derivatives in the tangential directions vanish.
The code outputs the evolution of the scalar field and, more importantly, the value of $\left | \gamma - 1 \right |$ at 8 kpc from the center---corresponding to the position of the Solar System in the Milky Way.

We confirm that the values used for technical parameters of our solver give a stable solution by running convergence tests. These are performed by increasing the resolution in factors of two (both temporal and spatial resolution separately) until the resulting scalar field at some later time $t_\mathrm{max}$ does not change significantly with resolution.

\begin{figure}
%\resizebox{\hsize}{!}{
\includegraphics[trim={0 0 10 0}, clip,width=\columnwidth]{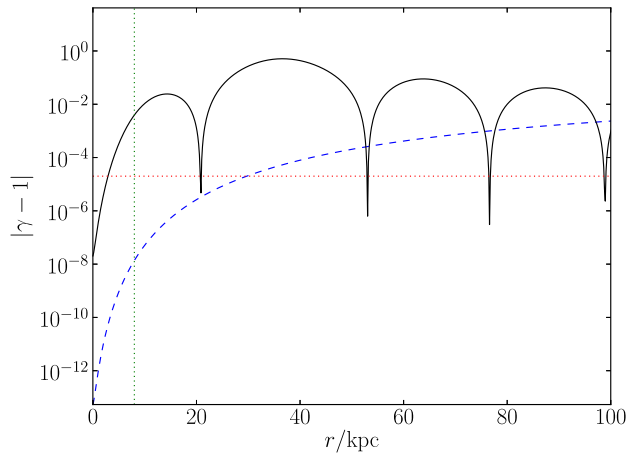}
%\includegraphics[trim={0 0 10 0}, clip]{onepanel-eps-converted-to.pdf}
%}
\caption{\label{fig:4panels} The PPN parameter $|\gamma - 1|$, plotted against distance from the center of the Galaxy. The curves show $|\gamma - 1|$ in the quasistatic case (blue dashed line), as well as after a scalar wave has entered the halo (black solid line). The vertical (green dotted) line indicates the position of the Solar System, and the horizontal (red dotted) line indicates the highest allowed value of $\gamma - 1$ in the Solar System from the Cassini experiment. When the wave enters the Milky Way, it increases the value of $|\gamma - 1|$ by several orders of magnitude.}
\end{figure}

%%%%%%%%%%%%%%%%%%%%%%%%%%%%%%%%%%%%%%%%%%%%%%%%%%%%%%%%%%%%%%%%%%%%%%%%%%%%%%%%
\emph{Results}.---Figure \ref{fig:4panels} shows an example of how the PPN parameter $\gamma$ changes when a wave enters the inner 100 kpc of the Milky Way.  The vertical line shows the position of the Solar System, which we assume to be 8 kpc from the Galactic center.  The modifications to gravity are initially screened very well in the regions around this position, with $|\gamma - 1| < 10^{-8}$ (blue dashed line). However, after the wave has arrived (black solid line), the scalar field is perturbed enough to breach the Solar System constraints, $|\gamma - 1| > 2\times 10^{-5}$. In other words, the screening mechanism breaks down under these circumstances. The wave in this particular simulation has an amplitude $A = 0.01$ and a frequency $\omega = 40 ~\mathrm{Myr}^{-1}$.  
The cusps are regions where the scalar field is zero, which exist since the wave oscillates both above and below $\chi = 0$. 

When measuring $\gamma$ arising from a single sinusoidal wave with low frequency, 
there is a possibility that the local wave is between two extrema at the time of measurement. 
This could render this kind of detection difficult for several thousand years. 
Nevertheless, given that various astrophysical events---such as supernovae---can generate waves,
the probability that one of the wavefronts would bring us away from the minima at the present time is not negligible.
%We expect waves arising naturally in other models to have higher frequencies, 
%and consequently these would be detectable within reasonable time.
%This effect should be further studied in the future, and can lead to interesting probes.

\begin{figure}[!h] %remove asterisk for one column figure.
%\resizebox{\hsize}{!}{
\includegraphics[trim={0 0 45 0}, clip,width=\columnwidth]{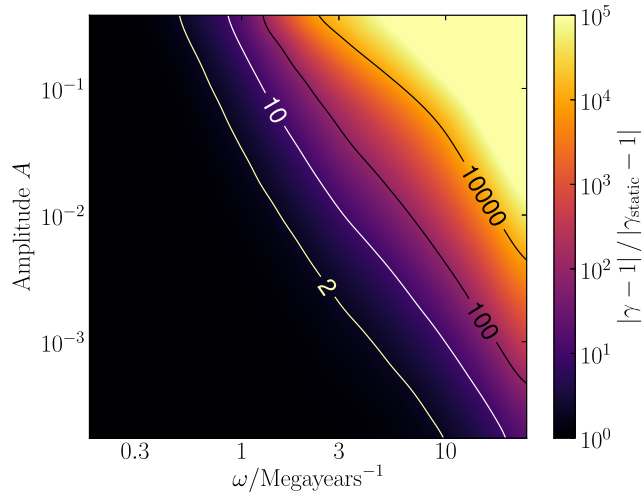}
%\includegraphics[trim={0 0 45 0}, clip]{gammalist-eps-converted-to.pdf}
%}
\caption{\label{fig:symgamma} Maximum increase in the PPN parameter $|\gamma - 1|$ due to incoming scalar field waves at the position of the Sun in the Galaxy (8 kpc from the center) as a function of amplitude and frequency of the incoming waves. The color indicates by which factor $|\gamma - 1|$ is increased when compared to the quasistatic case with no waves.}
\end{figure}

In order to investigate how our result depends on the frequency $\omega$ and amplitude $A$ of the waves, we simulate incoming waves with several values of these two parameters.  Figure \ref{fig:symgamma} shows the maximum growth of $|\gamma - 1|$ that we found at 8 kpc from the Galactic center. Brighter colors mean a larger increase of $|\gamma - 1|$ compared to the quasistatic approximation. The values of the frequency and amplitude that lie in the black region of the plot, give waves that do not significantly impact $\gamma$ compared to the quasistatic solution. Therefore, in this region of parameter space, the screening mechanism is efficient and hides the extra degree of freedom from gravity experiments.

%%%%%%%%%%%%%%%%%%%%%%%%%%%%%%%%%%%%%%%%%%%%%%%%%%%%%%%%%%%%%%%%%%%%%%%%%%%%%%%%

From Figure \ref{fig:symgamma}, it is possible to conclude that  higher frequencies and amplitudes for the incoming scalar waves give larger deviations from the general relativity result (i.e., $\gamma = 1$).  The limit where amplitude and frequency go to zero is equivalent to the quasistatic limit, where no waves are produced and their energy is zero.  As one goes into the high frequency and amplitude regime, the waves carry more energy, and therefore, the PPN parameter $\gamma$ starts deviating significantly from the quasistatic limit.  Note that, since in the symmetron model, the fifth force is $F_{\phi}\propto \nabla\phi^2/M^2$, these values can be immediately extrapolated to the impact of the waves on this quantity \citep{2010PhRvL.104w1301H}.

The  dependence of the $\gamma$ PPN parameter on the wave amplitude is straightforward to understand: When a wave propagates through the screened regions of the halo, a larger amplitude wave will lead to larger displacements of the field  from the screening value $\phi \approx 0$. Therefore,  $|\gamma - 1| \appropto \phi^2$ will increase accordingly. 

The frequency dependence of the $\gamma$ parameter is a consequence of the following: The effective potential of the symmetron grows steeper and narrower in high density areas. In other words, the mass of the field increases towards the center of the halo. Therefore,  it becomes more difficult to perturb the field away from the minimum, and a higher wave energy is needed to displace it. Specifically, if the energy of the external waves is small compared to the mass of the field, the field will not be perturbed and the $\gamma$ parameter will not be affected.

The results obtained in this Letter imply that if waves with sufficient amplitude or frequency can somehow be generated in a given model for modified gravity, they will have to be taken into account when constraining the model parameters. Cosmic tsunamis, resulting from extreme events, could even completely ruin the screening mechanisms in modified gravity by increasing the deviations from general relativity by several orders of magnitude compared to the quasistatic case.  A subject that that must be discussed now is the generation of such waves. 
Extreme events on small scales, such as collision of neutron stars, stellar, or super-massive black holes are obvious examples.  Generation of waves by pulsating stars are another possibility \cite{2011PhRvL.106y1101S}. 

In the specific case of the symmetron model, it is possible to obtain waves from events that occur on cosmological scales.  First, the symmetron model undergoes a phase transition when the density falls below a specific threshold.  This transition first occurs in voids when the expansion factor is close to $a_\mathrm{SSB}$ \cite{2013PhRvL.110p1101L, 2014PhRvD..89h4023L}.  When this happens, the scalar field receives a kick, which produces waves traveling from the center of the voids towards the dark matter halos.  By doing postprocessing of simulations presented in \cite{2016A&A...585A..37H}, we find that, in a symmetron model with slightly different parameters, the amplitude of cosmological waves is typically smaller that 0.1 and the associated frequencies are of the order of 1/Myr.  Note that these values depend on the model parameters and, hence, must be taken only as indicative. Scalar waves can also be created through the collapse of topological defects, which are known to exist in any model in which such phase transition occurs.  The energies associated with these kinds of waves are studied in \cite{2014PhRvD..90l4041L}.

\emph{Conclusions}.---Modifications to general relativity have long been studied, both when searching for the source of the accelerated expansion of the Universe, and to construct a UV complete theory of gravity. However, there are strong Solar System constraints on the deviation from Einstein's gravity when extending the theory by adding new degrees of freedom.  Thus, the viability of modified theories of gravity is strongly dependent on the existence of a screening mechanism that suppresses any extra degrees of freedom at these scales.  In this Letter, we show that waves propagating in an additional gravity degree of freedom, may significantly spoil the screening mechanism and, hence, jeopardize the viability of the given modified gravity theory.  Specifically, we show that waves in a given model can increase the amplitude of the fifth force and the post-Newtonian parameter $|\gamma - 1|$ by several orders of magnitude, rendering theories previously assumed to be viable unfeasible.

We reach our conclusions by performing numerical simulations of the propagation of waves through a Milky Way sized dark matter halo.  For a particular set of model parameters, we determine the importance of the amplitude and frequency of the incoming waves.  Increased amplitudes and frequencies (i.e., higher energy waves) lead to a greater impact on observables associated to the Solar System.

Our results were obtained in the context of a specific model of modified gravity, the symmetron model.  
We expect that they can be generalized to other models as well.  
For instance, including an extra disformal term in the coupling of the symmetron field 
can increase the amplitude of the oscillations of the scalar field in the center of the 
halos \cite{2016A&A...585A..37H}, thus, weakening the efficiency of the screening mechanism even further.  
Waves were also studied in a particular version of $f(R)$ theories \citep{2015JCAP...02..034B}.  
While the impact of these waves was found to be negligible in a cosmological context (i.e., structure formation), 
their effect in Solar System tests may still be detectable and help in further constraining the model.
To demonstrate that similar effects can be expected in other theories, 
we propose a simple calculation regarding a viable chameleon model \cite{ChameleonConstraints} ($n = 1,\, M = 10^{-3} M_\mathrm{Pl},\, \Lambda=2\,\mathrm{meV}$).
When a white dwarf explodes as a type Ia supernova, waves in the chameleon field will be measurable at several Mpc distance.
Full details on this calculation can be found in the Supplemental Material \cite{Supplemental}.

The applicability of the quasistatic approximation should be carefully analyzed when obtaining constraints for modified gravity theories from Solar System experiments.  Our results show that, in modified gravity, the Solar System---and indeed, the Galaxy---can not be studied in isolation; events that occur on cosmological scales might actually impact events that happen in the inner Solar System.  While our conclusions make it more difficult to build viable modified gravity theories based on screening mechanisms, the existence of nonstatic effects opens a completely new window for developing new tests of gravity.
 
\begin{acknowledgments}
We thank Chung-Chi Lee for useful discussion.
R.H. and D.F.M. acknowledge support from the Research Council of Norway, and the NOTUR facilities. C.L.L. acknowledges support from the STFC consolidated Grant No. ST/L00075X/1.
\end{acknowledgments}
\bibliography{references}

\begin{thebibliography}{21}%
\makeatletter
\providecommand \@ifxundefined [1]{%
 \@ifx{#1\undefined}
}%
\providecommand \@ifnum [1]{%
 \ifnum #1\expandafter \@firstoftwo
 \else \expandafter \@secondoftwo
 \fi
}%
\providecommand \@ifx [1]{%
 \ifx #1\expandafter \@firstoftwo
 \else \expandafter \@secondoftwo
 \fi
}%
\providecommand \natexlab [1]{#1}%
\providecommand \enquote  [1]{``#1''}%
\providecommand \bibnamefont  [1]{#1}%
\providecommand \bibfnamefont [1]{#1}%
\providecommand \citenamefont [1]{#1}%
\providecommand \href@noop [0]{\@secondoftwo}%
\providecommand \href [0]{\begingroup \@sanitize@url \@href}%
\providecommand \@href[1]{\@@startlink{#1}\@@href}%
\providecommand \@@href[1]{\endgroup#1\@@endlink}%
\providecommand \@sanitize@url [0]{\catcode `\\12\catcode `\$12\catcode
  `\&12\catcode `\#12\catcode `\^12\catcode `\_12\catcode `\%12\relax}%
\providecommand \@@startlink[1]{}%
\providecommand \@@endlink[0]{}%
\providecommand \url  [0]{\begingroup\@sanitize@url \@url }%
\providecommand \@url [1]{\endgroup\@href {#1}{\urlprefix }}%
\providecommand \urlprefix  [0]{URL }%
\providecommand \Eprint [0]{\href }%
\providecommand \doibase [0]{http://dx.doi.org/}%
\providecommand \selectlanguage [0]{\@gobble}%
\providecommand \bibinfo  [0]{\@secondoftwo}%
\providecommand \bibfield  [0]{\@secondoftwo}%
\providecommand \translation [1]{[#1]}%
\providecommand \BibitemOpen [0]{}%
\providecommand \bibitemStop [0]{}%
\providecommand \bibitemNoStop [0]{.\EOS\space}%
\providecommand \EOS [0]{\spacefactor3000\relax}%
\providecommand \BibitemShut  [1]{\csname bibitem#1\endcsname}%
\let\auto@bib@innerbib\@empty
%</preamble>
\bibitem [{\citenamefont {{Riess}}\ and\ \citenamefont
  {et~al.}(1998)}]{1998AJ....116.1009R}%
  \BibitemOpen
  \bibfield  {author} {\bibinfo {author} {\bibfnamefont {A.~G.}\ \bibnamefont
  {{Riess}}}\ and\ \bibinfo {author} {\bibnamefont {et~al.}},\ }\href {\doibase
  10.1086/300499} {\bibfield  {journal} {\bibinfo  {journal} {\aj}\ }\textbf
  {\bibinfo {volume} {116}},\ \bibinfo {pages} {1009} (\bibinfo {year}
  {1998})},\ \Eprint {http://arxiv.org/abs/astro-ph/9805201} {astro-ph/9805201}
  \BibitemShut {NoStop}%
\bibitem [{\citenamefont {{Weinberg}}(1989)}]{1989RvMP...61....1W}%
  \BibitemOpen
  \bibfield  {author} {\bibinfo {author} {\bibfnamefont {S.}~\bibnamefont
  {{Weinberg}}},\ }\href {\doibase 10.1103/RevModPhys.61.1} {\bibfield
  {journal} {\bibinfo  {journal} {Reviews of Modern Physics}\ }\textbf
  {\bibinfo {volume} {61}},\ \bibinfo {pages} {1} (\bibinfo {year}
  {1989})}\BibitemShut {NoStop}%
\bibitem [{\citenamefont {{Clifton}}\ \emph {et~al.}(2012)\citenamefont
  {{Clifton}}, \citenamefont {{Ferreira}}, \citenamefont {{Padilla}},\ and\
  \citenamefont {{Skordis}}}]{2012PhR...513....1C}%
  \BibitemOpen
  \bibfield  {author} {\bibinfo {author} {\bibfnamefont {T.}~\bibnamefont
  {{Clifton}}}, \bibinfo {author} {\bibfnamefont {P.~G.}\ \bibnamefont
  {{Ferreira}}}, \bibinfo {author} {\bibfnamefont {A.}~\bibnamefont
  {{Padilla}}}, \ and\ \bibinfo {author} {\bibfnamefont {C.}~\bibnamefont
  {{Skordis}}},\ }\href {\doibase 10.1016/j.physrep.2012.01.001} {\bibfield
  {journal} {\bibinfo  {journal} {\physrep}\ }\textbf {\bibinfo {volume}
  {513}},\ \bibinfo {pages} {1} (\bibinfo {year} {2012})},\ \Eprint
  {http://arxiv.org/abs/1106.2476} {arXiv:1106.2476 [astro-ph.CO]} \BibitemShut
  {NoStop}%
\bibitem [{\citenamefont {{Reynaud}}\ and\ \citenamefont
  {{Jaekel}}(2009)}]{2009aosp.conf..203R}%
  \BibitemOpen
  \bibfield  {author} {\bibinfo {author} {\bibfnamefont {S.}~\bibnamefont
  {{Reynaud}}}\ and\ \bibinfo {author} {\bibfnamefont {M.-T.}\ \bibnamefont
  {{Jaekel}}},\ }in\ \href@noop {} {\emph {\bibinfo {booktitle} {Atom Optics
  and Space Physics: Proceedings of the International School of Physics
  ``Enrico Fermi'', Course CLXVIII, Varenna on Lake Como, Villa Monastero,
  2007}}},\ \bibinfo {editor} {edited by\ \bibinfo {editor} {\bibfnamefont
  {E.}~\bibnamefont {{Arimondo}}}, \bibinfo {editor} {\bibfnamefont
  {W.}~\bibnamefont {{Ertmer}}}, \bibinfo {editor} {\bibfnamefont {W.~P.}\
  \bibnamefont {{Schleich}}}, \ and\ \bibinfo {editor} {\bibfnamefont {E.~M.}\
  \bibnamefont {{Rasel}}}}\ (\bibinfo  {publisher} {IOS Press, Amsterdam},\
  \bibinfo {year} {2009})\ p.\ \bibinfo {pages} {203},\ \Eprint
  {http://arxiv.org/abs/0801.3407} {arXiv:0801.3407 [gr-qc]} \BibitemShut
  {NoStop}%
\bibitem [{\citenamefont {Joyce}\ \emph {et~al.}(2014)\citenamefont {Joyce},
  \citenamefont {Jain}, \citenamefont {Khoury},\ and\ \citenamefont
  {Trodden}}]{Joyce:2014kja}%
  \BibitemOpen
  \bibfield  {author} {\bibinfo {author} {\bibfnamefont {A.}~\bibnamefont
  {Joyce}}, \bibinfo {author} {\bibfnamefont {B.}~\bibnamefont {Jain}},
  \bibinfo {author} {\bibfnamefont {J.}~\bibnamefont {Khoury}}, \ and\ \bibinfo
  {author} {\bibfnamefont {M.}~\bibnamefont {Trodden}},\ }\href@noop {} {\
  (\bibinfo {year} {2014})},\ \Eprint {http://arxiv.org/abs/1407.0059}
  {arXiv:1407.0059 [astro-ph.CO]} \BibitemShut {NoStop}%
%%CITATION = ARXIV:1407.0059;%%
\bibitem [{\citenamefont {{Llinares}}\ and\ \citenamefont
  {{Mota}}(2013)}]{2013PhRvL.110p1101L}%
  \BibitemOpen
  \bibfield  {author} {\bibinfo {author} {\bibfnamefont {C.}~\bibnamefont
  {{Llinares}}}\ and\ \bibinfo {author} {\bibfnamefont {D.~F.}\ \bibnamefont
  {{Mota}}},\ }\href {\doibase 10.1103/PhysRevLett.110.161101} {\bibfield
  {journal} {\bibinfo  {journal} {Physical Review Letters}\ }\textbf {\bibinfo
  {volume} {110}},\ \bibinfo {eid} {161101} (\bibinfo {year} {2013})},\ \Eprint
  {http://arxiv.org/abs/1302.1774} {arXiv:1302.1774} \BibitemShut {NoStop}%
\bibitem [{\citenamefont {Lindroos}\ \emph {et~al.}(2016)\citenamefont
  {Lindroos}, \citenamefont {Llinares},\ and\ \citenamefont {Mota}}]{Lindroos}%
  \BibitemOpen
  \bibfield  {author} {\bibinfo {author} {\bibfnamefont {J.~{\O{}}.}\
  \bibnamefont {Lindroos}}, \bibinfo {author} {\bibfnamefont {C.}~\bibnamefont
  {Llinares}}, \ and\ \bibinfo {author} {\bibfnamefont {D.~F.}\ \bibnamefont
  {Mota}},\ }\href {\doibase 10.1103/PhysRevD.93.044050} {\bibfield  {journal}
  {\bibinfo  {journal} {Phys. Rev. D}\ }\textbf {\bibinfo {volume} {93}},\
  \bibinfo {pages} {044050} (\bibinfo {year} {2016})},\ \Eprint
  {http://arxiv.org/abs/1512.00615} {arXiv:1512.00615 [gr-qc]} \BibitemShut
  {NoStop}%
\bibitem [{\citenamefont {{Hinterbichler}}\ and\ \citenamefont
  {{Khoury}}(2010)}]{2010PhRvL.104w1301H}%
  \BibitemOpen
  \bibfield  {author} {\bibinfo {author} {\bibfnamefont {K.}~\bibnamefont
  {{Hinterbichler}}}\ and\ \bibinfo {author} {\bibfnamefont {J.}~\bibnamefont
  {{Khoury}}},\ }\href {\doibase 10.1103/PhysRevLett.104.231301} {\bibfield
  {journal} {\bibinfo  {journal} {Physical Review Letters}\ }\textbf {\bibinfo
  {volume} {104}},\ \bibinfo {eid} {231301} (\bibinfo {year} {2010})},\ \Eprint
  {http://arxiv.org/abs/1001.4525} {arXiv:1001.4525 [hep-th]} \BibitemShut
  {NoStop}%
\bibitem [{\citenamefont {Khoury}\ and\ \citenamefont {Weltman}(2004)}]{cham}%
  \BibitemOpen
  \bibfield  {author} {\bibinfo {author} {\bibfnamefont {J.}~\bibnamefont
  {Khoury}}\ and\ \bibinfo {author} {\bibfnamefont {A.}~\bibnamefont
  {Weltman}},\ }\href {\doibase 10.1103/PhysRevLett.93.171104} {\bibfield
  {journal} {\bibinfo  {journal} {Phys.Rev.Lett.}\ }\textbf {\bibinfo {volume}
  {93}},\ \bibinfo {pages} {171104} (\bibinfo {year} {2004})},\ \Eprint
  {http://arxiv.org/abs/astro-ph/0309300} {arXiv:astro-ph/0309300 [astro-ph]}
  \BibitemShut {NoStop}%
%%CITATION = ASTRO-PH/0309300;%%
\bibitem [{\citenamefont {Koivisto}\ \emph {et~al.}(2012)\citenamefont
  {Koivisto}, \citenamefont {Mota},\ and\ \citenamefont
  {Zumalac{\'a}rregui}}]{dis4}%
  \BibitemOpen
  \bibfield  {author} {\bibinfo {author} {\bibfnamefont {T.~S.}\ \bibnamefont
  {Koivisto}}, \bibinfo {author} {\bibfnamefont {D.~F.}\ \bibnamefont {Mota}},
  \ and\ \bibinfo {author} {\bibfnamefont {M.}~\bibnamefont
  {Zumalac{\'a}rregui}},\ }\href {\doibase 10.1103/PhysRevLett.109.241102}
  {\bibfield  {journal} {\bibinfo  {journal} {Phys.Rev.Lett.}\ }\textbf
  {\bibinfo {volume} {109}},\ \bibinfo {pages} {241102} (\bibinfo {year}
  {2012})},\ \Eprint {http://arxiv.org/abs/1205.3167} {arXiv:1205.3167
  [astro-ph.CO]} \BibitemShut {NoStop}%
%%CITATION = ARXIV:1205.3167;%%
\bibitem [{\citenamefont {Burrage}\ and\ \citenamefont
  {Khoury}(2014)}]{Burrage}%
  \BibitemOpen
  \bibfield  {author} {\bibinfo {author} {\bibfnamefont {C.}~\bibnamefont
  {Burrage}}\ and\ \bibinfo {author} {\bibfnamefont {J.}~\bibnamefont
  {Khoury}},\ }\href {\doibase 10.1103/PhysRevD.90.024001} {\bibfield
  {journal} {\bibinfo  {journal} {Phys. Rev.}\ }\textbf {\bibinfo {volume}
  {D90}},\ \bibinfo {pages} {024001} (\bibinfo {year} {2014})},\ \Eprint
  {http://arxiv.org/abs/1403.6120} {arXiv:1403.6120 [hep-th]} \BibitemShut
  {NoStop}%
%%CITATION = ARXIV:1403.6120;%%
\bibitem [{\citenamefont {Babichev}\ \emph {et~al.}(2009)\citenamefont
  {Babichev}, \citenamefont {Deffayet},\ and\ \citenamefont {Ziour}}]{Babi}%
  \BibitemOpen
  \bibfield  {author} {\bibinfo {author} {\bibfnamefont {E.}~\bibnamefont
  {Babichev}}, \bibinfo {author} {\bibfnamefont {C.}~\bibnamefont {Deffayet}},
  \ and\ \bibinfo {author} {\bibfnamefont {R.}~\bibnamefont {Ziour}},\ }\href
  {\doibase 10.1142/S0218271809016107} {\bibfield  {journal} {\bibinfo
  {journal} {Int. J. Mod. Phys.}\ }\textbf {\bibinfo {volume} {D18}},\ \bibinfo
  {pages} {2147} (\bibinfo {year} {2009})},\ \Eprint
  {http://arxiv.org/abs/0905.2943} {arXiv:0905.2943 [hep-th]} \BibitemShut
  {NoStop}%
%%CITATION = ARXIV:0905.2943;%%
\bibitem [{\citenamefont {{Bertotti}}\ \emph {et~al.}(2003)\citenamefont
  {{Bertotti}}, \citenamefont {{Iess}},\ and\ \citenamefont
  {{Tortora}}}]{Cassini}%
  \BibitemOpen
  \bibfield  {author} {\bibinfo {author} {\bibfnamefont {B.}~\bibnamefont
  {{Bertotti}}}, \bibinfo {author} {\bibfnamefont {L.}~\bibnamefont {{Iess}}},
  \ and\ \bibinfo {author} {\bibfnamefont {P.}~\bibnamefont {{Tortora}}},\
  }\href {\doibase 10.1038/nature01997} {\bibfield  {journal} {\bibinfo
  {journal} {\nat}\ }\textbf {\bibinfo {volume} {425}},\ \bibinfo {pages} {374}
  (\bibinfo {year} {2003})}\BibitemShut {NoStop}%
\bibitem [{\citenamefont {{Llinares}}\ and\ \citenamefont
  {{Mota}}(2014)}]{2014PhRvD..89h4023L}%
  \BibitemOpen
  \bibfield  {author} {\bibinfo {author} {\bibfnamefont {C.}~\bibnamefont
  {{Llinares}}}\ and\ \bibinfo {author} {\bibfnamefont {D.~F.}\ \bibnamefont
  {{Mota}}},\ }\href {\doibase 10.1103/PhysRevD.89.084023} {\bibfield
  {journal} {\bibinfo  {journal} {\prd}\ }\textbf {\bibinfo {volume} {89}},\
  \bibinfo {eid} {084023} (\bibinfo {year} {2014})},\ \Eprint
  {http://arxiv.org/abs/1312.6016} {arXiv:1312.6016 [astro-ph.CO]} \BibitemShut
  {NoStop}%
\bibitem [{\citenamefont {{Silvestri}}(2011)}]{2011PhRvL.106y1101S}%
  \BibitemOpen
  \bibfield  {author} {\bibinfo {author} {\bibfnamefont {A.}~\bibnamefont
  {{Silvestri}}},\ }\href {\doibase 10.1103/PhysRevLett.106.251101} {\bibfield
  {journal} {\bibinfo  {journal} {Physical Review Letters}\ }\textbf {\bibinfo
  {volume} {106}},\ \bibinfo {eid} {251101} (\bibinfo {year} {2011})},\ \Eprint
  {http://arxiv.org/abs/1103.4013} {arXiv:1103.4013} \BibitemShut {NoStop}%
\bibitem [{\citenamefont {{Hagala}}\ \emph {et~al.}(2016)\citenamefont
  {{Hagala}}, \citenamefont {{Llinares}},\ and\ \citenamefont
  {{Mota}}}]{2016A&A...585A..37H}%
  \BibitemOpen
  \bibfield  {author} {\bibinfo {author} {\bibfnamefont {R.}~\bibnamefont
  {{Hagala}}}, \bibinfo {author} {\bibfnamefont {C.}~\bibnamefont
  {{Llinares}}}, \ and\ \bibinfo {author} {\bibfnamefont {D.~F.}\ \bibnamefont
  {{Mota}}},\ }\href {\doibase 10.1051/0004-6361/201526439} {\bibfield
  {journal} {\bibinfo  {journal} {\aap}\ }\textbf {\bibinfo {volume} {585}},\
  \bibinfo {eid} {A37} (\bibinfo {year} {2016})},\ \Eprint
  {http://arxiv.org/abs/1504.07142} {arXiv:1504.07142} \BibitemShut {NoStop}%
\bibitem [{\citenamefont {{Llinares}}\ and\ \citenamefont
  {{Pogosian}}(2014)}]{2014PhRvD..90l4041L}%
  \BibitemOpen
  \bibfield  {author} {\bibinfo {author} {\bibfnamefont {C.}~\bibnamefont
  {{Llinares}}}\ and\ \bibinfo {author} {\bibfnamefont {L.}~\bibnamefont
  {{Pogosian}}},\ }\href {\doibase 10.1103/PhysRevD.90.124041} {\bibfield
  {journal} {\bibinfo  {journal} {\prd}\ }\textbf {\bibinfo {volume} {90}},\
  \bibinfo {eid} {124041} (\bibinfo {year} {2014})},\ \Eprint
  {http://arxiv.org/abs/1410.2857} {arXiv:1410.2857} \BibitemShut {NoStop}%
\bibitem [{\citenamefont {{Bose}}\ \emph {et~al.}(2015)\citenamefont {{Bose}},
  \citenamefont {{Hellwing}},\ and\ \citenamefont
  {{Li}}}]{2015JCAP...02..034B}%
  \BibitemOpen
  \bibfield  {author} {\bibinfo {author} {\bibfnamefont {S.}~\bibnamefont
  {{Bose}}}, \bibinfo {author} {\bibfnamefont {W.~A.}\ \bibnamefont
  {{Hellwing}}}, \ and\ \bibinfo {author} {\bibfnamefont {B.}~\bibnamefont
  {{Li}}},\ }\href {\doibase 10.1088/1475-7516/2015/02/034} {\bibfield
  {journal} {\bibinfo  {journal} {\jcap}\ }\textbf {\bibinfo {volume} {2}},\
  \bibinfo {eid} {034} (\bibinfo {year} {2015})},\ \Eprint
  {http://arxiv.org/abs/1411.6128} {arXiv:1411.6128} \BibitemShut {NoStop}%
\bibitem [{\citenamefont {{Burrage}}\ and\ \citenamefont
  {{Sakstein}}(2016)}]{ChameleonConstraints}%
  \BibitemOpen
  \bibfield  {author} {\bibinfo {author} {\bibfnamefont {C.}~\bibnamefont
  {{Burrage}}}\ and\ \bibinfo {author} {\bibfnamefont {J.}~\bibnamefont
  {{Sakstein}}},\ }\href {http://stacks.iop.org/1475-7516/2016/i=11/a=045}
  {\bibfield  {journal} {\bibinfo  {journal} {Journal of Cosmology and
  Astroparticle Physics}\ }\textbf {\bibinfo {volume} {2016}},\ \bibinfo
  {pages} {045} (\bibinfo {year} {2016})},\ \Eprint
  {http://arxiv.org/abs/1609.01192} {arXiv:1609.01192} \BibitemShut {NoStop}%
\bibitem [{Sup()}]{Supplemental}%
  \BibitemOpen
  \href@noop {} {{\bibinfo {title} {See supplemental material
  in the Appendix}}\ }\BibitemShut {NoStop}%
\bibitem [{Note1()}]{Note1}%
  \BibitemOpen
  \bibinfo {note} {From the top left panel of Figure 3 in \cite
  {ChameleonConstraints}, one can check that $M_\protect \mathrm {c}/M_\protect
  \mathrm {pl} = 10^{-3}$ and $\Lambda = 2 \times 10^{-4}\protect \tmspace
  +\thinmuskip {.1667em}\protect \mathrm {eV}$ is outside of the excluded
  regions. In our notation, $\beta = M_\protect \mathrm {pl}/M_\protect \mathrm
  {c} = 10^3$ and $M = \Lambda = 0.2\protect \tmspace +\thinmuskip
  {.1667em}\protect \mathrm {meV}$}\BibitemShut {NoStop}%
\end{thebibliography}%
%%%%%%%%%%%%%%%%%%%%%%%%%%%%%%%%%%%%%%%%%%%%%%%%%%%%%%%%%%%%%%%%%%%%%%%%%%%%%%%%
%APPENDIX
%
\newpage
\subsection*{Supplemental Material: Chameleon Waves from Supernova Type Ia}

In this supplementary material we do an order-of-magnitude
estimate of the effects of Chameleon scalar field waves generated by a type Ia supernova.
We will first find the energy that can be released into scalar field waves in such an event,
and then consider the amplitude of scalar waves with this energy. Finally,
we will predict the measurable impact on $\left| \gamma - 1 \right|$ from waves with this amplitude.
All calculations are initially performed with natural units $\hbar=c=1$, with preferred units of mass.
These will then be translated to lengths or times as needed.

Let us regard a simple $n=1$ Chameleon field with effective potential\begin{equation}
V_{\mathrm{eff}}=\rho\left(1+\frac{\beta}{M_{\mathrm{Pl}}}\phi\right)+\frac{M^{5}}{\phi}.\label{eq:Veff}\end{equation}
\\
For such a model, the scalar field value at the minimum of the
potential and the effective mass will be given by

\begin{eqnarray}
\phi_{\mathrm{min}} & = & \left(\frac{M^{5}M_{\mathrm{Pl}}}{\rho\beta}\right)^{1/2},\label{eq:phimin}\\
m_{\mathrm{eff}} & = & \left(\frac{4\rho^{3}\beta^{3}}{M_{\mathrm{Pl}}^{3}M^{5}}\right)^{1/4}.\end{eqnarray}

To be specific, we set \begin{eqnarray*}
M & = & 0.2\,\mathrm{meV}\approx10^{-31}M_{\mathrm{Pl}},\\
\beta & = & 10^{3},\end{eqnarray*}
\\
in agreement with all current observational and experimental constraints
\cite{ChameleonConstraints}\footnote{From the top left panel of Figure 3 in 
\cite{ChameleonConstraints}, one can check that $M_\mathrm{c}/M_\mathrm{pl} = 10^{-3}$ 
and $\Lambda = 2 \times 10^{-4}\,\mathrm{eV}$ is outside of the excluded regions. 
In our notation, $\beta = M_\mathrm{pl}/M_\mathrm{c} = 10^3$
and $M = \Lambda = 0.2\,\mathrm{meV}$}. 
This choice corresponds to a Chameleon range
in the background density of the universe of $\lambda_{c}=1/m_{\mathrm{eff}}=0.04$
parsec. 

For the supernova, we will assume the simplest possible model: a white
dwarf -- located in or close to our Milky Way -- with radius $2\times10^{6}$
m, and mass $1.4M_{\odot}$; it has a density of about $\rho_{\mathrm{WD}}=10^{37}\rho_{0}$.
This white dwarf is completely destroyed over some short time, spreading
the matter evenly over several parsec such that $\rho_{\mathrm{end}}\ll\rho_{\mathrm{WD}}$.
Inside the white dwarf, the Chameleon field range will be sub-micrometre
scale, so we can safely assume $\phi=\phi_{\mathrm{min}}$ throughout
the white dwarf. 

The proposal is that there is a significant energy difference for the scalar field
when the effective potential goes from a density of $\rho_{\mathrm{WD}}$
to $\rho_{\mathrm{end}}$. This energy difference will be released (at least partially)
in the form of scalar field waves. Assuming the field follows the
minimum of $V_{\mathrm{eff}}$ as the density changes, one can insert
the expression for the field value \eqref{eq:phimin} into the effective potential equation
\eqref{eq:Veff}.\begin{equation}
V_{\mathrm{eff}}=\rho+2\left(\frac{M^{5}\beta\rho}{M_{\mathrm{Pl}}}\right)^{1/2}.\label{eq:Veff-1}\end{equation}
\\
The change in the effective potential when the density goes from $\rho_{\mathrm{WD}}$
to $\rho_{\mathrm{end}}$ can then be found from\begin{equation}
\Delta V_{\mathrm{eff}}=\rho_{\mathrm{WD}}-\rho_{\mathrm{end}}+
2\left(\frac{M^{5}\beta}{M_{\mathrm{Pl}}}\right)^{1/2}\left( \rho_{\mathrm{WD}}^{1/2} - \rho_{\mathrm{end}}^{1/2} \right).\label{eq:deltaVeff}\end{equation}

Using the conversion factors\begin{eqnarray*}
\rho_{0} & = & 3\Omega_{0}H_{0}^{2}M_{\mathrm{Pl}}^{2}\approx H_{0}^{2}M_{\mathrm{Pl}}^{2},\\
H_{0} & = & 2\times10^{-4}\,\mathrm{Mpc}^{-1}\approx6\times10^{-61}M_{\mathrm{Pl}},\end{eqnarray*}
\\
one finds the energy density expected at the source of the supernova,
\begin{equation}
\Delta V_{\mathrm{eff}}\approx\rho_{\mathrm{WD}}\approx3.6\times10^{-84}M_{\mathrm{Pl}}^{4}.\end{equation}
\\
The other terms of equation \eqref{eq:deltaVeff} are many orders
of magnitude smaller and can be safely neglected for this purpose.
Assuming no Hubble friction, this energy will be diluted with $1/r^{2}$
as the waves travel radially outward, until they reach the Solar System. 

Now we need to estimate the amplitude of oscillations of the field
resulting from this release. Close to the sun -- where Cassini measured
$\gamma$ \cite{Cassini} -- the average density is around $\rho_{\odot}\approx10^{29}\rho_{0}$.
If the field is at rest at the bottom of the effective potential,
it will have a field value dictated by equation \eqref{eq:phimin}.
For small $\phi$, the Chameleon will have a $\gamma$ parameter following
$\left|\gamma-1\right|\propto\phi$. Inserting
numbers, we find\begin{eqnarray*}
\phi_{\mathrm{min},\odot} & = & \left(2.7\times10^{-67}M_{\mathrm{Pl}}^{2}\right)^{1/2}\\
 & = & 5.2\times10^{-34}M_{\mathrm{Pl}}.\end{eqnarray*}
\\
\textbf{To increase $\left|\gamma-1\right|$ by a factor of 10},
we would need oscillations within the potential with enough energy
to lift $\phi$ from $\phi_{\mathrm{min},\odot}$ to $10\times\phi_{\mathrm{min},\odot}$
given a constant density $\rho_{\odot}$. To find numbers for the energy density needed,
we can calculate the difference in the value of the effective potential $V_\mathrm{eff}$
when going from from $\phi_{\mathrm{min},\odot}$ to $10\times\phi_{\mathrm{min},\odot}$.
\begin{eqnarray}
\Delta V_{\mathrm{eff}} & = & V_{\mathrm{eff}}\left(\phi=10\phi_{\mathrm{min},\odot}\right)-V_{\mathrm{eff}}\left(\phi=\phi_{\mathrm{min},\odot}\right)\\
 & = & 9\rho_{\odot}\frac{\beta}{M_{\mathrm{Pl}}}\phi_{\mathrm{min},\odot}-\frac{9M^{5}}{10\phi_{\mathrm{min},\odot}},\nonumber \\
 & = & \frac{81}{10}\left(\frac{\rho_{\odot}\beta M^{5}}{M_{\mathrm{Pl}}}\right)^{1/2},\nonumber \\
 & \approx & 1.5\times10^{-121}M_{\mathrm{Pl}}^{4}.\end{eqnarray}
\\
When comparing this to the scalar field energy released in the
supernova (equation \ref{eq:deltaVeff}), one can find that even if
less than 1\% of the energy is released into waves (the rest can be
transferred to the nearby matter through the fifth force), one should
be able to measure a 10-fold increase in $\left|\gamma-1\right|$
as far as 10 Mpc away from the supernova -- a distance within which
there are several hundreds of galaxies. This serves as 
proof that there exists phenomena that can release measurable scalar
field waves in viable Chameleon models as well as the symmetron model
presented in the Letter.

On a final note: if we are at a minimum and not at a maximum
of a wave, it could significantly lower the value of $\left|\gamma-1\right|$. How
long time will it take for the measurable maximum to arrive? To find the frequency
of the waves, let us model the effective potential around $\phi_{\mathrm{min}}$
to second order in $\delta\phi$, we find\begin{equation}
V_{\mathrm{eff}}\approx V_{\min}+M^{5}\left(\frac{\rho\beta}{M^{5}M_{\mathrm{Pl}}}\right)^{3/2}\left(\delta\phi\right)^{2},\end{equation}
\\
a simple harmonic oscillator with natural frequency\begin{equation}
\omega_{0}=\sqrt{2M^{5}}\left(\frac{\rho\beta}{M^{5}M_{\mathrm{Pl}}}\right)^{3/4}.\end{equation}
\\
Considering the energy scales we have been using in the Letter and in this reply,  $\delta\phi$
is not small enough for this approximation to hold perfectly, but the order
of magnitude should nevertheless give us a useful estimate. Inserting
the average density of the universe $\rho_{0}$, and the chosen Chameleon
parameters, we find\begin{eqnarray*}
\omega_{0} & = & \sqrt{2\times10^{-155}M_{\mathrm{Pl}}^{5}}\left(3.6\times10^{37}M_{\mathrm{Pl}}^{-2}\right)^{3/4},\\
\omega_{0} & = & 6.6\times10^{-50}M_{\mathrm{Pl}}\approx2\times10^{-7}\,\mathrm{Hz},\end{eqnarray*}
\\
in the order of a couple oscillations per year. 
Such a frequency is viable to observe in e.g. the rate of change of $\gamma$ over time.

\end{document}